\documentclass[%
 reprint, superscriptaddress,
 amsmath,
 amssymb,
 aps,
nolongbibliography,noeprint,pra,aps,graphics,a4paper,twocolumn
]{revtex4-2}
\usepackage{graphicx}
\usepackage{dcolumn}
\usepackage{float}
\usepackage{bm}
\usepackage[margin=25mm]{geometry}
\usepackage{graphicx}
\usepackage{siunitx}
\usepackage{physics}
\usepackage{amsmath,bbm}
\usepackage{braket}
\usepackage{ragged2e}
\usepackage{tikz}
\usepackage{pgfplots}
\usepackage{tikzscale}
\usepackage{siunitx}
\usepackage{hyperref}
\usepackage{gensymb}
\usepackage[]{subcaption}
\begin{document}
\preprint{APS/123-QED}
\newcommand{\xone}{\vec{x}_1}
\newcommand{\xtwo}{\vec{x}_2}
\newcommand{\pone}{\vec{p}_1}
\newcommand{\ptwo}{\vec{p}_2}
\newcommand{\xonehat}{\hat{\vec{x}}_1}
\newcommand{\xtwohat}{\hat{\vec{x}}_2}
\newcommand{\ponehat}{\hat{\vec{p}}_1}
\newcommand{\ptwohat}{\hat{\vec{p}}_2}
\newcommand{\dxone}{\Delta \hat{ \vec{x}}_1}
\newcommand{\dxtwo}{\Delta \hat{ \vec{x}}_2}
\newcommand{\be}{\begin{equation}}
\newcommand{\ee}{\end{equation}}
\newcommand{\ba}{\begin{array}}
\newcommand{\ea}{\end{array}}
\newcommand{\bqa}{\begin{eqnarray}}
\newcommand{\eqa}{\end{eqnarray}}
\newcommand{\Inv}{{\cal I}}
\newcommand{\Itwo}{\Gamma}
\newcommand{\Ione}{\vec{W}}
\newcommand{\nion}{{N}}
\newcommand{\ddim}{{d}}
\newcommand{\IO}{\theta}
\newcommand{\unitary}{U}
\newcommand{\Htrapfreq}{M}
\newcommand{\overbar}[1]{\mkern 1.5mu\overline{\mkern-1.5mu#1\mkern-1.5mu}\mkern 1.5mu}
\newcommand{\Htrapfreqbar}{\mathcal{M}}
\newcommand{\shifted}{\mathcal{N}}
\newcommand{\com}{A}
\newcommand{\Hforce}{\vec{F}}
\newcommand{\Htrapforcebar}{{\vec{\mathcal{F}}}}
\newcommand{\traj}{\vec{L}}
\newcommand{\flo}[1]{{\color[rgb]{0,0.3,0.6}{#1}}}
\newcommand{\hil}[1]{{\color[rgb]{0.7,0.2,0.2}{#1}}}
\newcommand{\selwyn}[1]{{\color[rgb]{0,0.6,0.3}{#1}}}
\newcommand{\symplectic}{{\cal S}}
\newcommand{\duration}{T}
\newcommand{\coupling}{D}
\newcommand{\couplingbar}{\mathcal{D}}
\newcommand{\antisyminteg}{{\cal J}} 
\newcommand{\pos}{R}
\newcommand{\Hcenter}{\vec{C}}
\title{Quantum invariant-based control of interacting trapped ions}
\author{Selwyn Simsek}%
\affiliation{
Blackett Laboratory, Imperial College London, South Kensington Campus, London SW7 2AZ, UK
}%

\author{Florian Mintert}%
\affiliation{
Blackett Laboratory, Imperial College London, South Kensington Campus, London SW7 2AZ, UK
}%
\date{\today}
\begin{abstract}
Invariant-based inverse engineering
is an elegant approach to quantum control
with corresponding experimental implementations that perform tasks 
with applications in quantum information processing such as shuttling trapped ions.
We build on recent work to generalise invariant-based inverse engineering to control two coupled harmonic oscillators in any number of spatial dimensions. This may be used to perform experimentally relevant tasks such as separation of trapped ions, which is demonstrated numerically, achieving transfer fidelities of over 96\% as well as low motional number excitations.
\end{abstract}
\maketitle

\section{Introduction}
Trapped ions are among the most developed platforms for quantum information processing.
Both single-qubit quantum gates and entangling gates can be implemented with high fidelity \cite{Weidt, Ballance2016, webbResilientEntanglingGates2018},
and there exists the prospect of realising a scalable architecture with microfabricated, structured traps.
The resulting quantum charge-coupled device 
can trap a large number of ions in various trapping zones,
and shuttling ions between such zones results in a high interconnectivity of all the trapped ions~\cite{Kielpinski2002,Lekitsch2017,webberEfficientQubitRouting2020b,pinoDemonstrationTrappedionQuantum2021}.

While the development of logical gates for trapped ions is fairly advanced with many schemes for fast and noise-resilient gate operations~\cite{timoneyErrorresistantSinglequbitGates2008,schaferFastQuantumLogic2018, webbResilientEntanglingGates2018}, 
shuttling of trapped ions is a much less mature field.
Because of limited coherence times, it is important to separate an ion from one chain of interacting ions and to shuttle it towards another chain as fast as possible~\cite{Lekitsch2017}. 
Since the realization of logical gates requires the motion 
of trapped ions to be close to their ground state~\cite{webbResilientEntanglingGates2018},
it is equally important that motional excitations of the ions at the end of any shuttling protocol are kept to a minimum.

Ideally, a shuttling protocol would thus transfer ions that initially occupy their motional ground state to the ground state of a final trapping potential.
Such a task fits exactly into the setting of adiabatic control, but the necessarily slow dynamics conflicts greatly with the requirement to have fast shuttling protocols.
Invariant-based inverse engineering is well suited to construct shuttling protocols that ensure ground state to ground state transfer without the requirement of slow dynamics.
In particular for harmonic potentials, the framework of quantum invariants is rather well developed, and the harmonic approximation is excellent for trapped ions.
Invariant-based inverse engineering has thus become a frequently used technique in the context of trapped ions with many conceptual developments~\cite{torronteguiFastAtomicTransport2011,Palmero2013,Furst2013,luFastShuttlingTrapped2014,luOptimalTransportTwo2015,palmeroFastExpansionsCompressions2015,palmeroFastSeparationTwo2015,palmeroShortcutsAdiabaticityIon2016,lizuainDynamicalNormalModes2017,levyNoiseResistantQuantum2018,luFastShuttlingParticle2018} and experimental implementations with atoms and trapped ions~\cite{nessRealisticShortcutsAdiabaticity2018,kaushalShuttlingbasedTrappedionQuantum2020}.
While most approaches to quantum dynamics would focus on solving the dynamics induced by a given Hamiltonian, invariant-based approaches rely on an Ansatz for the dynamics and aim at constructing a Hamiltonian that gives rise to this desired dynamics.
The challenge lies in ensuring that the Ansatz results in a Hamiltonian that can be realized in practice.
While an Ansatz resulting in a Hamiltonian that is quadratic in position and momentum operators may be found reasonably straightforwardly, only the trapping potential of trapped ions can be modified in practice. The kinetic energy, however, is determined by the mass of the ions and any Ansatz resulting in a Hamiltonian with a different kinetic energy does not help to find a practically realizable shuttling protocol.

There exists a class of invariants that result in Hamiltonians with a given kinetic energy term that can be used to find time-dependent trapping potentials that realise ground state to ground state shuttling protocols for individual trapped ions \cite{simsek_quantum_2020}.
Generalizing such approaches to interacting trapped ions, however, brings a crucial further difficulty: while the actual trapping potential of the trapped ions can be chosen at will, the Coulomb interaction between trapped ions cannot be modified in any way.
Any approach that relies on the ability to choose the full potential term of the system Hamiltonian would thus not result in practically realizable shuttling protocols.
Incorporating the constraints imposed by interactions is possible if the dynamics of each ion can be restricted to one single motional degree of freedom \cite{tobalinaInvariantbasedInverseEngineering2020}, but many operations required for quantum information procession, such as swapping two ions \cite{splattDeterministicReorderingOf40Ca2009,kaufmannFastIonSwapping2017} do not admit a one-dimensional approximation.

The main contribution of this paper is the derivation of a quantum invariant that results in a Hamiltonian with given kinetic energy and pairwise interaction between the ions without a restriction on the number of motional degrees of freedom of any ion.
While a rigorous basis for invariant-based inverse engineering is given for invariants with a non-degenerate ground state, the present invariant has a degenerate spectrum.
Despite this caveat, however, this invariant can be used to devise shuttling protocols that are implemented in terms of time-dependent trapping potentials only,
as demonstrated with the example of separating two initially interacting ions.
Even though these degeneracies result in slight imperfections, the framework results in accurate shuttling protocols well beyond the adiabatic approximation.

\section{Quantum invariants}
\label{sec:background}
A quantum invariant $\Inv$ is a Hermitian operator that satisfies the equation of motion
\be \label{eqn:invariant-eom}
\frac{\partial \Inv(t)}{\partial t}=i[\Inv(t),H(t)],
\ee
with the generally time-dependent system Hamiltonian $H(t)$.
Crucially, the instantaneous eigenstates of an invariant may be chosen to be solutions of the time-dependent Schr\"odinger equation with the Hamiltonian $H(t)$ \cite{lewisExactQuantumTheory1969}.

If the invariant $\Inv$ has a non-degenerate ground state, and $\Inv$ commutes with the Hamiltonian $H(t)$ at initial and final times, which is to say that $[\Inv(0),H(0)]=0$ and $[\Inv(\duration),H(\duration)]=0$,
then the ground state of $\Inv$ is also an eigenstate of $H(t)$ at $t=0$ and $t=T$.
One has thus found a solution of the time-dependent Schr\"odinger equation that yields the evolution of an eigenstate of the initial Hamiltonian $H(0)$ towards an eigenstate of the final Hamiltonian $H(t)$.

With invariant-based inverse engineering, one would start out with an Ansatz for the time-dependent invariant $\Inv(t)$, and then construct a Hamiltonian such that the equation of motion Eq.~\eqref{eqn:invariant-eom} is satisfied.
The crucial difficulty lies in finding an invariant that gives rise to a Hamiltonian that can be experimentally realised.

For the translational degrees of freedom of a single trapped ion, for example,
the trapping potential  can be tuned in terms of voltages applied to the trap electrodes,
but the kinetic energy term dependent on momentum and mass of the ion is an intrinsic system property that can not be modified in practice.
It is absolutely essential that the Hamiltonian resulting from an invariant has exactly this kinetic energy term.

Going from a single trapped ion or several non-interacting trapped ions to the problem of interacting trapped ions, yields the Coulomb interaction as another component to the Hamiltonian that can not be modified.

It is thus essential that the Hamiltonian resulting from an invariant is of the form with the intrinsic kinetic energy of the trapped ions and the Coulomb interaction as given by the position of the ions.
Only single-ion potential terms may appear as quantities that need to be chosen such that Eq.~\eqref{eqn:invariant-eom} is indeed satisfied.

Realistically, this is achievable only within some approximation, such as the Gaussian approximation in which the Hamiltonian is approximated in a second order Taylor expansion around the expected position of the ions.
For the Coulomb interaction, this implies the approximation 
\begin{align}
\frac{1}{\lvert \xonehat-\xtwohat \rvert} \simeq &
\frac{1}{|\vec r|} - \frac{\vec\delta_{\hat x}\vec{r}}{|\vec r|^3} - \frac{\vec\delta_{\hat x}\vec\delta_{\hat x}}{2 |\vec r|^3}+ \frac{3}{2}\frac{(\vec\delta_{\hat x}\vec{r})^2}{|\vec r|^5}\ ,
\label{eqn:coulomb-expansion}
\end{align}
with
\be
\vec r=\xone - \xtwo \hspace{.5cm}\mbox{and}\hspace{.5cm} \vec\delta_{\hat x}=\xonehat -\xtwohat -\vec r
\ee
defined in terms of the expectation values $\xone = \langle  \xonehat \rangle$ and $\xtwo = \langle \xtwohat \rangle$ the (vectorial) positions of the two ions.

The full Hamiltonian for $\nion$ ions within the Gaussian approximation is thus of the form
\begin{align}
  H &= \sum_{i=1}^{\nion}\frac{\vec{\hat p}_i^2}{2 m_i} + \frac{1}{2} m_i \vec{\hat x}_i^T \Htrapfreqbar_i \vec{\hat x}_i- \Htrapforcebar_i \cdot \vec{\hat x}_i
  \nonumber\\
  &
   + \sum_{i< j}^{\nion}\vec{\hat x}_i^T \coupling(\vec x_i- \vec x_j)\ \vec{\hat x}_j\ .
    \label{eqn:coupled-hamiltonian}
\end{align}
It contains an interaction matrix $\coupling(\vec x_i- \vec x_j)$ with
\be
\coupling(\vec r)= C\left(\frac { \mathbbm{1}}{|\vec r|^3} - \frac{3\vec r \vec r^T}{|\vec r|^5}\right)\ ,
\ee
that depends on the spatial separation of the Gaussian wave-packets of pairs of ions.
The real-symmetric matrices $\Htrapfreqbar_i$ and real vectors $\Htrapforcebar_i$ that characterize the harmonic potential experienced by the individual ions, have contributions of both the actual trapping potential and the Coulomb interaction with the other ions.
Since any contribution from the Coulomb interaction to the single-ion terms of the Hamiltonian can be compensated by the trapping potential,
the matrices $\Htrapfreqbar_i$ and vectors $\Htrapforcebar_i$ are thus taken as freely tuneable objects,
whereas $\coupling(\vec r)$ can not be modified at all through changes in the trapping potential.

The ground state of any such quadratic Hamiltonian is Gaussian. 
It is thus characterized in terms of the expectation value $Z=\langle X\rangle$ of the $2\nion\ddim$-dimensional vector
\be
X=\begin{pmatrix} \xonehat , \xtwohat , \hdots , \hat x_{\nion}, \ponehat , \ptwohat  , \hdots, \hat p_{\nion} \end{pmatrix}.
\ee
of phase space operators, and the corresponding covariance matrix $\Sigma$ with elements
\be
\Sigma_{ij}=\frac{1}{2}\langle X_iX_j+X_jX_i\rangle-\langle X_i\rangle\langle X_j\rangle\ .
\ee
If the Hamiltonian changes in time while remaining quadratic, any initially Gaussian state remains Gaussian,
but the phase space vector $Z$ and the covariance matrix become time-dependent.
This is independent of whether the change in the Hamiltonian is adiabatic or not, and the vector $Z$ satisfies the classial Hamiltonian equations of motion and can be understood as the classical phase space trajectory of the ions.

\section{A quantum invariant for interacting particles}
\label{sec:new-invariant}

The present invariant 
\be
\Inv = \frac{1}{2}(X-Z)^T \Gamma (X-Z)
\label{eqn:invariant-ansatz}
\ee
is a regular quadratic function of the vector $X$ of phase space operators and the phase-space trajectory $Z$ of the ions.
With the form of a general quadratic operator, $\Inv$ is an invariant ({\it i.e.} satisfies Eq.~\eqref{eqn:invariant-eom}) for a general quadratic Hamiltonian
\be
H = \frac{1}{2}X^T\Omega X+V \cdot X
\ee
if and only if the equations of motion
\begin{align}
\dot{\Gamma} &= \Omega \symplectic \Gamma - \Gamma \symplectic \Omega, \label{eqn:quadratic-eom} \\
\dot{Z} &= \symplectic V+ \symplectic \Omega Z\label{eqn:linear-eom}
\end{align}
for $\Gamma$ and $Z$
with the $2\nion\ddim$-dimensional symplectic matrix 
\be
\symplectic=
\begin{pmatrix}
\mathbbm{O} & \mathbbm{1} \\
- \mathbbm{1} & \mathbbm{O}
\end{pmatrix}
\ee
are satisfied.

\subsection{Quadratic part}

The specific features that make $\Inv$ satisfy the desired properties are encoded in the real-symmetric matrix
\be
\Gamma = \Re \begin{pmatrix}
G^\dagger G
\end{pmatrix}\ , \label{eqn:gamma-definition}
\ee
defined in terms of the $\ddim\times\nion\ddim$ dimensional matrix
\begin{align}
G&=(S\otimes\mathbbm{1})\begin{pmatrix} {Y}_1 &
\hdots & Y_\nion & m_1\dot{Y}_1 &
 \hdots &  m_\nion\dot{Y}_\nion \end{pmatrix}
\label{eq:defG}\\
&=\begin{pmatrix} m_1\dot{Y}_1 &
 \hdots &  m_\nion\dot{Y}_\nion& -{Y}_1 &
  \hdots &-Y_\nion \end{pmatrix},
\label{eq:defG}
\end{align}
comprised of complex, square matrices $Y_i$ each referring to one individual ion.

In order to find the equations of motion for the matrices $Y_i$ that result from Eq.~\eqref{eqn:quadratic-eom},
it is helpful to notice that, in the case of interacting ions with the Hamiltonian given in Eq.~\eqref{eqn:coupled-hamiltonian},
the matrix $\Omega$ has the explicit form
\be
\Omega=\begin{pmatrix}
\Omega_{xx} & \mathbbm{O} \\
\mathbbm{O} & \Omega_{pp}
\end{pmatrix}\ ,
\ee
with
\be
\Omega_{pp}=
\begin{pmatrix}
\frac{\mathbbm{1}}{m_1} & 0 & 0 & 0 &\hdots & 0\\
0 & \frac{\mathbbm{1}}{m_2} & 0 & 0 &\hdots & 0\\
0 & 0 & \frac{\mathbbm{1}}{m_3} & 0 &\hdots & 0\\
\vdots & \vdots & \vdots & & & \vdots\\
0 & 0 & 0 & \hdots & 0 & \frac{\mathbbm{1}}{m_\nion}
\label{eqn:quadratic-hamiltonian}
\end{pmatrix},
\ee
for the kinetic energy, and
\be
\Omega_{xx}=
\begin{pmatrix}
m_1 \Htrapfreqbar_1 & \coupling[1,2] & \coupling[1,3] & \hdots & \coupling[1,\nion]\\
\coupling[1,2] & m_2 \Htrapfreqbar_2 & \coupling[2,3] & \hdots & \coupling[2,\nion]\\
\coupling[1,3] & \coupling[2,3] & m_3 \Htrapfreqbar_3 & \hdots & \coupling[3,\nion]\\
\vdots & \vdots & \vdots & & \vdots\\
\coupling[1,\nion] & \coupling[2,\nion] & \coupling[3,\nion] &\hdots & m_\nion \Htrapfreqbar_\nion
\end{pmatrix},
\ee
for the potential energy, where $D[i,j]$ is a short hand notation for $D(\vec x_i- \vec x_j)$.

With this explicit form of $\Omega$, the equation of motion for $G$
resulting from Eq.~\eqref{eqn:quadratic-eom} reads
\be
Re \left( G^\dagger \left( \dot{G} + G \symplectic \Omega \right) + \left( \dot{G} + G \symplectic \Omega \right)^\dagger G \right) = 0\ .
\label{eq:eqomG}
\ee
With the specific choice of $G$ given in Eq.~\eqref{eq:defG}, one obtains the explicit form 
\be
\dot{G} + G \symplectic \Omega  =
\begin{pmatrix}
Q_1&Q_2&\hdots &Q_\nion &\mathbbm{O}&\hdots&\mathbbm{O}
\end{pmatrix}.
\ee
with
\be
Q_i=m_i\left(  Y_i \Htrapfreqbar_i  + \ddot{Y}_i\right)+ \sum_{j\neq i}Y_j\coupling[i,j] \ . 
\ee
The equation of motion (Eq.~\eqref{eq:eqomG}) to be solved, is satisfied if all the $Q_i$ vanish, which is the case if the differential equations
\be
\ddot{Y}_i + Y_i \Htrapfreqbar_i   = -\sum_{j\neq i}\frac{Y_j\coupling[i,j]}{m_i} \ , \label{eqn:coupled-q1}
\ee
for the complex matrices $Y_i$ are satisfied.

\subsection{Construction of the trapping potential}

With the parametrization of the invariant $\Inv$ given in Eqs.\eqref{eqn:invariant-ansatz}, \eqref{eqn:gamma-definition} and \eqref{eq:defG}
the ion dynamics is defined in terms of the phase space trajectory $Z$ and the time-dependent matrices $Y_i$.
Any choice for those quantities completely determines the right-hand-side of Eq.~\eqref{eqn:coupled-q1}.
The matrices $\Htrapfreqbar_i$ that characterize the quadratic component of the trapping potential for the individual ions are thus determined by

\be
\Htrapfreqbar_i = -Y_i^{-1}\left(\ddot{Y}_i+\sum_{j\neq i}\frac{Y_j\coupling[i,j]}{m_i}\right)\ .
\label{eqn:Mi}
\ee

This is the desired prescription of how to chose the trapping potential, given in terms of the matrices $\Htrapfreqbar_i$,
such that the equation of motion Eq.~\eqref{eqn:invariant-eom} for the invariant is satisfied.
Crucially, ensuring the validity of the equation of motion is what
results in the determination of the regular trapping potential, and indeed no modification of the interaction is necessary.

There is, however, an aspect that requires some further care.
The matrices $\Htrapfreqbar_i$ given by Eq.~\eqref{eqn:Mi} are not necessarily symmetric and real as they would need to be in order to specify a harmonic trapping potential.
It is thus essential to ensure that this property will be given while choosing the matrices $Y_i$.
Similar to the one-dimensional case~\cite{ramos-prietoErmakovLewisInvariantTwo2020},
 the problem can be simplified sufficiently further for symmetric trapping potentials.
As discussed in more detail in Sec.~\ref{sec:symmetric} this will indeed result in the construction of real and symmetric matrices $\Htrapfreqbar_i$ that correspond to actual trapping potentials.

\section{Inverse engineering the invariant}
\label{sec:symmetric}

In the case of two ions, there is only one interaction term $\coupling(\vec x_1-\vec x_2)$.
For a symmetric trap configuration with $\Htrapfreqbar_1=\Htrapfreqbar_2=\Htrapfreqbar$,
the two ($i=1,2$) equations of motion in Eq.~\eqref{eqn:coupled-q1} are thus of exactly the same form.
One can therefore work with one single matrix
$P=\sqrt{m_1}Y_1=\sqrt{m_2}Y_2$ to define an invariant, as was considered recently \cite{ramos-prietoErmakovLewisInvariantTwo2020} in a one-dimensional context.
 
The equation of motion (Eq.~\eqref{eqn:coupled-q1}) for $P$ then reduces to
\be
\ddot{P} + P
\left( 
\Htrapfreqbar+\frac{\coupling}{\sqrt{m_1m_2}}
\right)=0\ ,
\label{eq:YM}
\ee
where the argument $[i,j]$ of the interaction matrix $\coupling$ is dropped.
This is exactly of the form of the special case of Eq.~\eqref{eqn:coupled-q1}) for non-interacting ions with the terms $\coupling(\vec x_i- \vec x_j)$ vanishing.

The problem is thus reduced to the case of a single trapped ion with the only difference that the dynamics depends on the modified quadratic component
\be
\tilde\Htrapfreqbar=\Htrapfreqbar+\frac{\coupling}{\sqrt{m_1m_2}}
\ee
of the potential, instead of the component $\Htrapfreqbar$ as given in the system Hamiltonian in Eq.~\eqref{eqn:coupled-hamiltonian}.

It is thus sufficient to resort to existing solutions~\cite{simsek_quantum_2020} for single ions that ensure that $\tilde \Htrapfreqbar$ and thus also $\Htrapfreqbar$ is real symmetric.

In this context it is helpful to express $P$ in a polar decomposition $P=\unitary\pos$
with a unitary $U$ and a positive semi-definite matrix $R$ that satisfies the initial condition $\dot\pos(0)=0$.
The matrix $\Gamma$ of the invariant $\Inv$ can then be determined by the choice of $R(t)$ instead of the choice of $P(t)$,
and the quadratic component $\Htrapfreqbar$ of the trapping potential resulting from a choice of $R(t)$ is determined by the relation 

\be
\{\pos^2,\tilde\Htrapfreqbar\}=2[\dot\pos,\pos]_{\com}-2\pos\com^2\pos-\{\ddot \pos, \pos\}\ ,
\label{eq:eomR}
\ee
with the generalized commutator $[\alpha,\beta]_\gamma=\alpha\gamma\beta-\beta\gamma\alpha$,
and the matrices $\com$ and $\antisyminteg$ satisfying
\be
\com=i\pos^{-2}+\frac{1}{2}[\pos^{-1},\dot \pos]+\frac{1}{2}\pos^{-1}{\antisyminteg}\pos^{-1}\ ,
\label{eq:A}
\ee
and
\be
\{{\antisyminteg},\pos^{-2}\} = [\dot \pos,\pos^{-1} ]+ [\pos,\pos^{-2}]_{\dot\pos}\ .
\label{eqn:j-definition}\\
\ee
The unitary $\unitary$ in the polar decomposition is specified to satisfy $\com=\unitary^\dagger \dot{\unitary}$.
Even though the matrix $\tilde\Htrapfreqbar$ enters Eq.~\eqref{eq:eomR} in terms of an anti-commutator with $\pos^2$,
Eqs.~\eqref{eq:eomR} to \eqref{eqn:j-definition} determine $\Htrapfreqbar$ uniquely, since the anti-commutator as linear map is indeed invertible.
The resulting matrix $\Htrapfreqbar$ is provably real and symmetric~\cite{simsek_quantum_2020}.

In order to ensure that the time-dependent Hamiltonian realises ground state to ground state transfer, it is important that $\Inv$ and $H$ commute at the initial time $t=0$ and the final time $t=\duration$ of the protocol.
Requiring this commutativity
is equivalent to requiring that the time-derivative $\dot\Inv$
vanishes because of the equation of motion $\dot\Inv=i[\Inv,H]$ of quantum invariants (Eq.\eqref{eqn:invariant-eom}).

Since the invariant $\Inv$ is parametrized in terms of the phase-space trajectory $Z(t)$ of the ions and the positive semi-definite matrix $R(t)$,
this implies that the time derivatives $\dot Z$ and $\dot\pos$ need to vanish at $t=0$ and $t=\duration$.

\begin{figure*}
  \begin{subfigure}[b]{0.66\columnwidth}
    \scalebox{1}[1]{\includegraphics[width=\linewidth,clip]{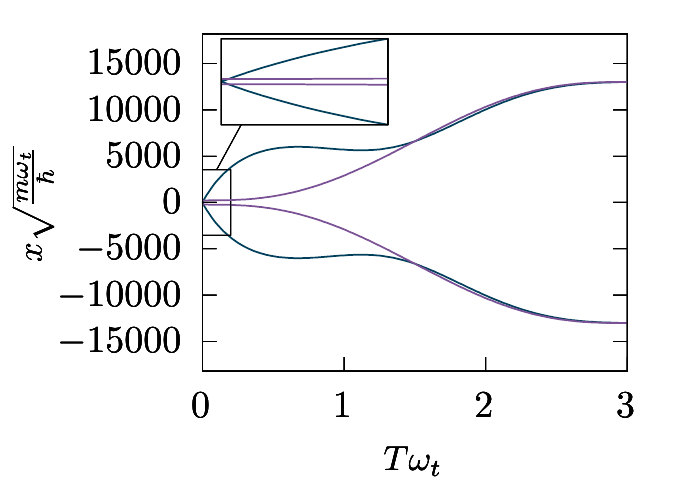}}
    \caption{$\duration=3 \omega_t ^{-1}$}
  \end{subfigure}%
  \begin{subfigure}[b]{0.66\columnwidth}
    \scalebox{1}[1]{\includegraphics[width=\linewidth,clip]{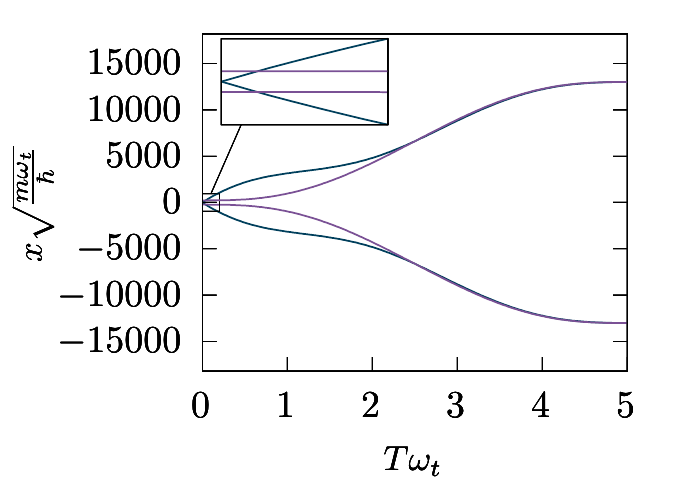}}
    \caption{$\duration=5 \omega_t ^{-1}$}
  \end{subfigure}%
  \begin{subfigure}[b]{0.66\columnwidth}
    \scalebox{1}[1]{\includegraphics[width=\linewidth,clip]{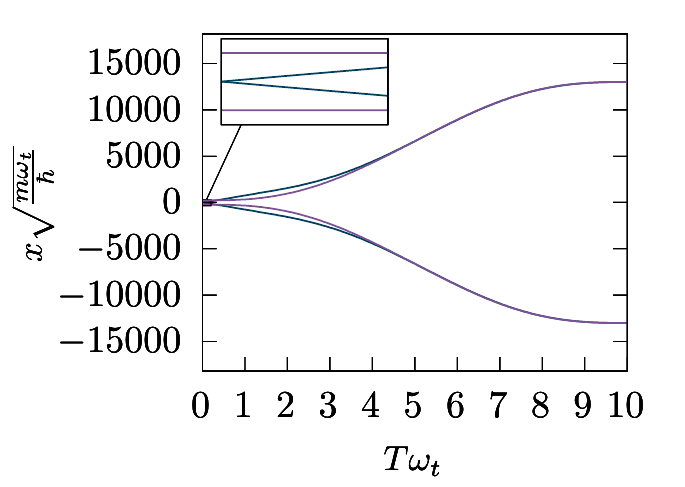}}
    \caption{$\duration=10 \omega_t ^{-1}$}
  \end{subfigure}%
  \caption{Separation of two ions into a T-junction. Inset (a) depicts a fast, ostensibly diabatic protocol, whereas inset (c) depicts a close-to-adiabatic protocol;
  inset (b) depicts an intermediate protocol.
  The dynamics of the trap centers are depicted by blue lines, and the dynamics of the ions are depicted by purple lines.
  The dynamics of trap centers and ions deviate increasingly with increasing diabaticity, but the shuttling protocols ensure that the ion trajectories end up at the classical equilibrium positions.
  Due to the finite Coulomb interaction, the initial equilibrium positions of the ions does not coincide with the trap center as highlighted in the insets.}
  \label{fig:separation-example}
\end{figure*}

The condition that $\dot Z$ vanishes asserts that only phase space trajectories may be chosen such that the velocities of the ions and the classical forces vanish, which is exactly the classical boundary conditions.
The condition that $\dot\pos$ vanishes requires a bit more care.
Since $\Gamma$ is parameterized in terms of $R$ and $\dot R$, the time-derivative $\dot\Gamma$ is a function of $R$ and its first two derivatives.
It is thus natural to require that $\dot R$ and $\ddot R$ vanish at $t=0$ and $t=\duration$.
At instances at which $\dot R$ and $\ddot R$ vanish,
Eqs.~\eqref{eq:eomR}, \eqref{eq:A}  and \eqref{eqn:j-definition} reduce to
\be
\{\pos^2,\tilde\Htrapfreqbar\}=-2\pos\com^2\pos\ ,\
\com=i\pos^{-2}\ \mbox{and}\ {\antisyminteg}=0\ .
\ee
Vanishing derivatives of $R$ thus result in the boundary condition
\be
\pos(t) =\left(\Htrapfreqbar(t) + \frac{\coupling}{\sqrt{m_1 m_2}}\right)^{-\frac{1}{4}} 
\ee
to be satisfied for $t=0$ and $t=\duration$.

\section{Ion separation} 
\label{sec:examples}

The capability of the present invariant for the construction of shuttling protocols can be demonstrated with the example of separating two ions in one spatial direction while they are moving along a second spatial direction.
This example will also demonstrate that despite the degenerate spectrum of the invariant, the accuracy of state transfer is substantially higher than that required to improve shuttling experimentally.

Initially, the ions are trapped in the same potential well centered at $[0,0]$ and separated only by their Coulomb repulsion.
At the final time $\duration$, the ions are trapped in two distinct potential wells centered at $[x,y]$ and $[-x,y]$ with a mutual separation $d=2x$.

The confinement
\be
\Htrapfreq(0)=\begin{pmatrix}
\omega_{t}^2 & 0 \\
0 & \omega_{r}^2
\end{pmatrix}\ , \ \
\Htrapfreq(\duration)=\begin{pmatrix}
\omega_{r}^2 & 0 \\
0 & \omega_{t}^2
\end{pmatrix}\ ,
\label{eq:boundary_confine}
\ee
with $\omega_r\gg\omega_t$
of the initial and final trapping potential is chosen such that the direction of strong confinement rotates during the separation process,
as would be the case in a T-junction \cite{Hensinger2006}.

The trapping potential together with the Coulomb interaction determines the initial and final positions $\vec x_1$ and $\vec x_2$ of the two ions.
The actual trajectories can be chosen freely
as long the the required boundary conditions are satisfied.
One may thus take the trajectories
\be
\vec x_i(\tau)=(1-p(\tau))\vec x_i(0)+p(\tau)\vec x(\duration)\ ,
\ee
with
\be
p(\tau) = 10\tau^3 - 15\tau^4 + 6\tau^5\ , \mbox{ and }\ \tau=t/T\ .
\ee
\begin{figure}
    \includegraphics[width=\linewidth]
    {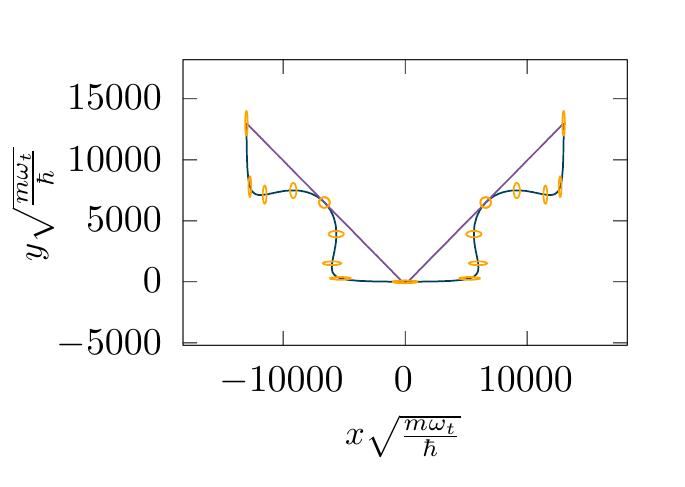}
  \caption{
    Two ions, initially trapped together in a T-junction, are separated into two distinct trapping locations at an angle of 90 degrees, over a duration $\duration=3\omega_t^{-1}$.
    The trap centre trajectories, plotted here in purple and orange, have superimposed on them level sets of the trapping frequencies in light blue and dark blue. The ions themselves travel on straight line trajectories shown in green and purple.
  }
   \label{fig:covariances}
\end{figure}

The time-evolving confinement $\Htrapfreq(t)$ is determined in terms of the positive semi-definite matrix $\pos(t)$ via Eq.~\eqref{eq:eomR}.
The boundary conditions
\be
\pos(0)=\left(
\Htrapfreqbar(0)+\frac{D}{m}\right)^{-\frac{1}{4}}
\ee
and
\bqa
\pos(\duration)&=&\left(
\Htrapfreqbar(\duration)+\frac{D}{m}\right)^{-\frac{1}{4}}\\
&\simeq&
\begin{pmatrix}
\omega_{r}^{-\frac{1}{2}} & 0 \\
0 & \omega_{t}^{-\frac{1}{2}}
\end{pmatrix}
\eqa
together with the choice $\pos(\tau)=(1-p(\tau))\pos(0)+p(\tau)\pos(\duration)$
are consistent with Eq.~\eqref{eq:boundary_confine}.

The following numerical examples are based on the explicit choices
$\omega_r=10\omega_t$, $\omega_t= 2 \pi \times 1\SI{}{\mega\hertz}$, and where $m$ is chosen to be the mass of a ytterbium-171 ion.
The ions are separated to a distance $d=200 \SI{}{\micro\meter} $, at which point the effect of the Coulomb interaction becomes negligible and the ions are effectively non-interacting.

Fig.~\ref{fig:separation-example} depicts three examples of shuttling protocols of different duration, with a fast ($\duration=3 \omega_t ^{-1}$) protocol in inset $a)$, a protocol of intermediate duration ($\duration=5 \omega_t ^{-1}$) in inset $b)$ and a close-to-adiabatic protocol ($\duration=10 \omega_t ^{-1}$) in inset $c)$.

The trajectories of the centers of the trapping potentials are depicted by blue lines, and the trajectories of the ions are depicted by purple lines.
Each inset depicts a zoom into the initial part of the trajectories. 
In this part, the positions of the ions do not coincide with the centers of the trapping potential because of their mutual Coulomb interaction.
At the final positions, however, the Coulomb interaction is negligibly small, and there is no discernible difference between center of trapping potential and ion position.
With increasing adiabaticity, the trajectories of the ions get closer to the trajectories of the trapping potential, and, in particular, in inset $a)$ the ions get substantially displaced from their equilibrium positions.

\begin{figure*}
  \begin{subfigure}[b]{0.5\columnwidth}
    {\includegraphics[width=\linewidth,clip]{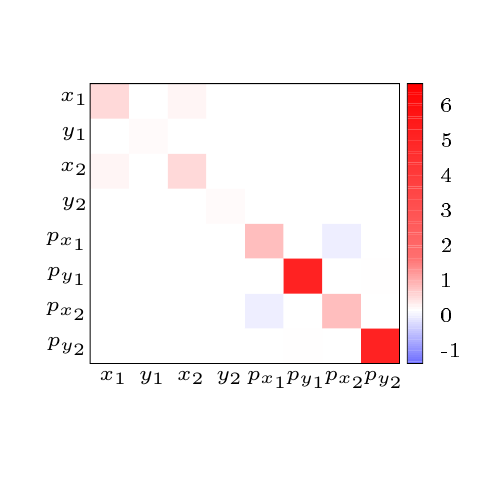}}
    \caption{$t=0$}
  \end{subfigure}%
  \begin{subfigure}[b]{0.5\columnwidth}
    {\includegraphics[width=\linewidth,clip]{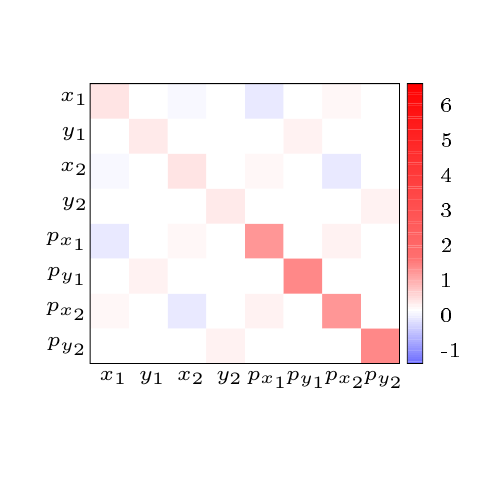}}
    \caption{$t=\frac{1}{2}\duration$}
  \end{subfigure}%
  \begin{subfigure}[b]{0.5\columnwidth}
    {\includegraphics[width=\linewidth,clip]{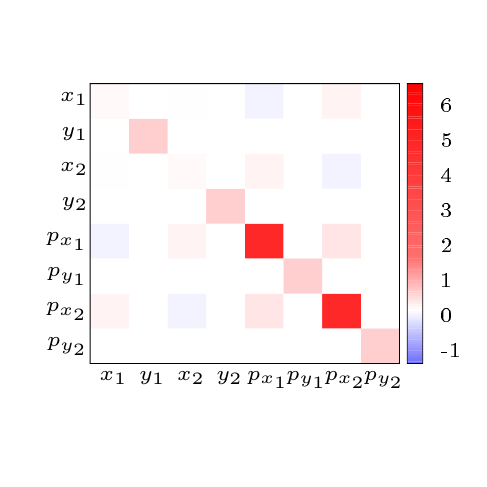}}
    \caption{$t=\duration$}
  \end{subfigure}%
  \begin{subfigure}[b]{0.5\columnwidth}
    {\includegraphics[width=\linewidth,clip]{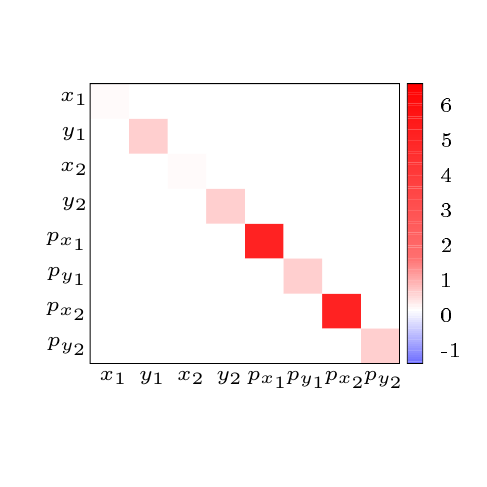}}
    \caption{Ground state at $t=\duration$}
  \end{subfigure}%
  \caption{
  Covariance matrices at different instances of the fast shuttling protocol of duration $T=3 \omega_t^{-1}$.
  The individual matrix elements are displayed in color-scale in harmonic oscillator units:
  position-position correlations have the unit $m\hbar\omega_t$, position-position correlations have the unit $\hbar/(m\omega_t)$, and position-momentum correlations have the unit $\hbar$.
  The starting state of the system depicted in (a) contains some position-position and momentum-momentum correlations between the ions resulting from the sizeable Coulomb interaction.
  In particular the position-momentum correlations undergo some transient dynamics (inset b) which are typical of states far from the ground state of a harmonic potential.
  Most of these correlations are negligibly small at the end of the protocol (inset c),
  but the correlation of the $p_x$-momenta between the two ions remains finite,
  whereas it vanishes in the correlation matrix of the ground state of the final trapping potential.
  The complete time-dependence of the $p_{x_1}$-$p_{x_1}$ correlation and the $p_{x_1}$-$p_{x_2}$-correlation is depicted in Fig.~\ref{fig:continuous-time-plots}.}
\label{fig:covariance-heatmaps-2}
\end{figure*}

Fig.~\ref{fig:covariances} depicts the dynamics of the confinement of the trapping potential in terms of equipotential lines at different positions in the plane.
The initial and final potentials are rotated with respect to each other by $90\degree$ as specified by the boundary condtions in Eq.~\eqref{eq:boundary_confine}.
In addition to the rotation of the trapping potential, the potential also becomes isotropic during the dynamics before reaching the final an-isotropic potential specified by the boundary conditions.
This deformation of the confining potential results in the desired dynamics of the covariances of the ions shown in Fig.~\ref{fig:covariance-heatmaps-2} that ensures that the ions end up close to their motional ground state.

\begin{figure}
\includegraphics[width=\linewidth,clip]{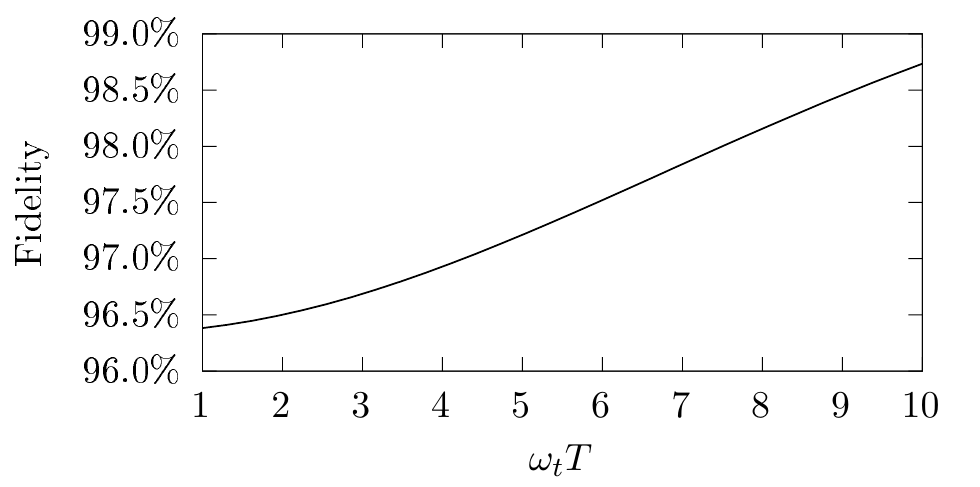}
    \caption{Fidelity of the ion separation protocol as a function of the protocol duration $\duration$.
    For adiabatic dynamics ($\duration\to\infty$), the adiabatic theorem guarantees a perfect fidelity, but for fast protocols there is no such guarantee.
    The present protocol realise high fidelities even for fast protocols, far outside the range of adiabatic dynamics.}
\label{fig:fidelity}
\end{figure}

Since the matrix $G$ defined in Eq.~\eqref{eq:defG} is rectangular, the matrix $\Gamma$ in Eq.~\eqref{eqn:gamma-definition} has a non-vanishing null-space, resulting in a degenerate spectrum of the invariant $\Inv$ (Eq.~\eqref{eqn:invariant-ansatz}).
As mentioned above, the rigorous footing for the working of the ground-state-to-ground-state is thus not given, but as shown in Fig.~\ref{fig:fidelity} the  trapping potentials derived from this invariant result in transfers with very high fidelities.
Fig.~\ref{fig:fidelity} depicts the state-fidelity of the above protocol as function of the duration $\duration$.
In the limit of slow protocols ($\duration\to\infty$) the success of the shuttling protocol is guaranteed due to adiabatic dynamics,
and indeed the fidelities approach the ideal value of unit fidelity with increasing duration $T$.
Even for fast protocols, however, the fidelities are very high and clearly exceed a value of $96\%$.
Since the degeneracy is attributed to the quadratic part of the invariant, it is guaranteed that the ions trajectories satisfy the desired properties; the expected positions and momenta of the ion do thus match the desired values at the end of the trapping potential, but the covariances of the ions do not necessarily end up in exactly their desired values.

Fig.~\ref{fig:covariance-heatmaps-2} depicts instances of the covariance matrix for the fast shuttling protocol of duration $\duration=3 \omega_t^{-1}$ depicted in inset $a)$ of Fig.~\ref{fig:separation-example}.
The initial covariance ($t=0$) is shown in inset $a)$; the covariance matrix in the middle of the shuttling protocol ($t=\frac{\duration}{2}$) is shown in inset $b)$; the final ($t=\duration$) covariance matrix and the desired, final covariance matrix are shown in insets $c)$ and $d)$.

The uncertainties in position and momentum reflect the properties of initial and final trapping potential: 
at $t=\duration$ the spatial uncertainty in the $x$ direction is larger than in the $y$ direction, whereas the opposite is the case at $t=\duration$,
and a similar pattern exists for the uncertainty in the momenta.
The initial covariance matrix (inset $a$) has a sizeable $x_1$-$x_2$-covariance and $p_{x_1}$-$p_{x_2}$-covariance as a result  of the strong Coulomb interaction between the ions.
In the ideal final state (inset $d$) all inter-ion covariances are negligibly small.
In the actual final covariance matrix (inset $c)$), also the $x_1$-$x_2$ correlation has become negligible,
but the magnitude of the $p_{x_1}$-$p_{x_2}$ covariance has increased as compared to its initial value.
The covariance matrix at $t=\frac{\duration}{2}$ (inset $b$) has several finite elements that are vanishing or negligibly small at both $t=0$ and $t=\duration$, which highlights the non-monotonic dynamics of the covariances.

\begin{figure*}
  \begin{subfigure}[b]{\columnwidth}
    {\includegraphics[width=\linewidth,clip]{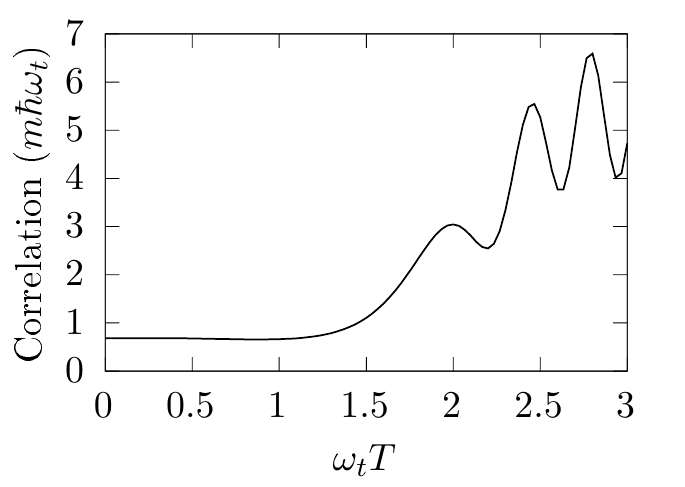}}
    \caption{$\Sigma_{p_{x_1} p_{x_1}}$}
  \end{subfigure}%
  \begin{subfigure}[b]{\columnwidth}
    {\includegraphics[width=\linewidth,clip]{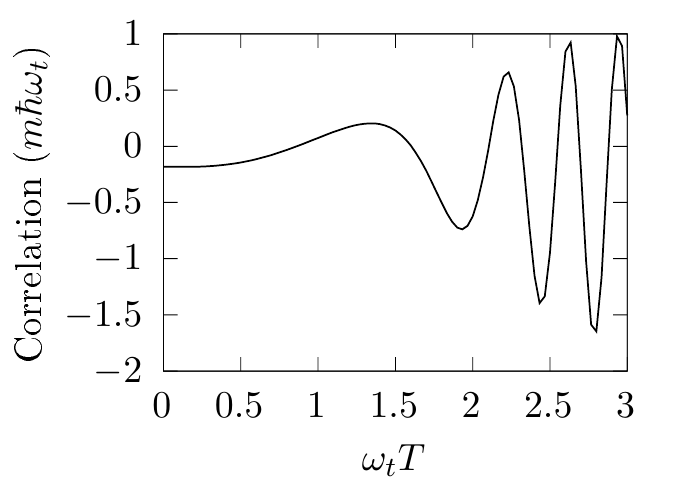}}
    \caption{$\Sigma_{p_{x_1} p_{x_2}}$}
  \end{subfigure}%
  \caption{
  Time-dependence of the $p_{x_1}$-$p_{x_1}$ correlation (inset a) and the $p_{x_1}$-$p_{x_2}$ correlation (inset b) for the fast shuttling protocol of duration $T=3 \omega_t^{-1}$.
  In particular due to the rotation between initial and final trapping potential the initial and final value of the $p_{x_1}$-$p_{x_1}$ correlation are different.
  The shuttling protocol realises a non-monotonic dynamics that ensures that the final value of this correlation matches the desired value.
  The desired value of the $p_{x_1}$-$p_{x_2}$ correlation is negligibly small because of the weak interaction between the separated ions.
  The shuttling protocol ensures that the actual final value of the $p_{x_1}$-$p_{x_2}$ correlation is smaller than the values adopted during the dynamics, but due to the invariant's degeneracy there is a small mismatch between the actual and the desired value.}
\label{fig:continuous-time-plots}
\end{figure*}

Fig.~\ref{fig:continuous-time-plots} depicts the actual time-dependence of two elements of the covariance matrix.
Inset $a)$ depicts the time-dependence of the $p_{x_1}$-$p_{x_1}$ correlation.
This correlation is non-vanishing at both the start and the end of the protocol. Over time, it develops some oscillations, and eventually settles on a much higher value than it had at the start of the protocol. Nevertheless, some deviation with the expected value is observed at the end of the protocol.
Inset $b)$ depicts the time-dependence of the $p_{x_1}$-$p_{x_2}$-covariance. This value is expected to become negligible at the end of the separation procedure, as may be seen by inspecting Fig.~\ref{fig:covariance-heatmaps-2}.
Physically, this is due to the fact at large ion separations the Coulomb potential is negligible and one may expect there to be no correlations between the two ions in their ground state. Nevertheless, as one can see from inset $b)$, some oscillations build up over time and at the end of the protocol, the covariance does not attain a value close to 0 as it should.

The deviations of the covariances from their desired values implies that the ions do not end up exactly in their motional ground state, but that there are some motional excitations.
The dominant contributions are occupations of the states $\ket{\nion_{x_1}\nion_{y_1}\nion_{x_2}\nion_{y_1}}=\ket{1010}$ ($1.57\%$), $\ket{2000}$ ($0.79\%$) and $\ket{0020}$ ($0.79\%$), with all occupation numbers of higher states becoming negligible.
These excitations are substantially below what is required for the realization of quantum gates \cite{Haddadfarshi2016,sametiStrongcouplingQuantumLogic2021}.

\section{Conclusion}

With the rapid advances in controlling trapped ions and current activities towards scaling up hardware for quantum information processing,
shuttling of trapped ions requires practical control techniques.
The invariant developed here offers an efficient approach for the construction of ground-state-to-ground-state shuttling protocols that work far outside the validity of the adiabatic approximation,
and the ability to go beyond the one-dimensional approximation enables control also in challenging trap geometries as encountered for example around junctions \cite{blakestadHighFidelityTransport2009,wrightReliableTransportMicrofabricated2013}.

Similar to experiments on control of motional quantum states, extensions to the theoretical framework of suitable invariants are also a great challenge.
Dynamical mode expansions \cite{palmeroFastTransportMixedspecies2014,palmeroFastSeparationTwo2015,lizuainDynamicalNormalModes2017} offer an alternative to the wave-packet dynamics used in this paper;
they offer the prospect to control mixed ion species~\cite{palmeroFastTransportMixedspecies2014} and have proven helpful for separation of trapped ions \cite{palmeroFastSeparationTwo2015}.
Even though symmetries -- as assumed here for the trap geometry -- can be very helpful for explicit solutions to control problems, a framework to support laboratory experiments also needs the capability to address problems without simplifying symmetries.
Finding invariants that satisfy all these wishes has proven very difficult, but the rapid advances in artificial intelligence over the last years gives hope for accelerated progress in the development of practically useful quantum invariants.

\section{Acknowledgements}
We are indebted to stimulating discussions with Adam Callison, Modesto Orozco Ruiz, Sam Hile, David Bretaud, Alex Owens, Pedro Taylor-Burdett, Sebastian Weidt and Winnie Hensinger.
This work was supported through a studentship
in the Centre for Doctoral Training in Controlled Quantum Dynamics at Imperial College London funded
by EPSRC(EP/L016524/1).

\bibliography{bibliography}

\end{document}